\documentstyle[11pt,paspconf]{article}

\begin{document}

\title{Quasars}

\author{Patrick S. Osmer}

\affil{Department of Astronomy, The Ohio State University, 174 W.
18th Ave., Columbus, OH 43210, USA, posmer@astronomy.ohio-state.edu}

\begin{abstract}

I review recent results for quasars and discuss how they are related to 
activity in galaxies.  Topics included are studies of quasar host
galaxies with HST; searches for quasars in the Hubble Deep Field;
evolution of the quasar luminosity function; news highlights from
astro-ph; and current observational research problems
and their relation to theoretical work.  

\end{abstract}

\section{Introduction}

Our meeting occurs 35 years after the discovery of quasars, a discovery that 
transformed our concepts of active galactic nuclei (AGN), 
even though the connection between quasars and AGN was not clear at the
time.  We now consider quasars as the most luminous class of AGNs.  Their great
luminosity, which can be more than 1000 times that of an $L^*$ galaxy, is  part
of their mystery on the one hand, while on the other hand it enables us to 
observe them at the greatest distances and earliest epochs at which they occur in the universe.

One of the great values of this symposium is that it brings together people from all
fields of AGN research and provides us with an opportunity to take a fresh look
at the state of the field and the key research problems.

In this talk I will cover some of the highlights of quasar history as they apply to our
topic and review the main properties of quasars as we define them today.  I will discuss
recent results in quasar research, especially those that bear on the relation of quasars to
activity in galaxies.  I will also describe some current research problems and consider 
future opportunities  for the field that will be provided by large telescopes and the
large quasar surveys that are under way.

\section{History}

While the discovery of quasars in 1963 (Schmidt 1963)
is well known\footnote{The 1st Texas Symposium
on Relativistic Astrophysics (Robinson, Schild, \& Schucking 1965)
still makes excellent reading about that feverish first year
of work.}, I would like to mention that it was preceded in 1958, forty years ago, by
a key paper by Burbidge at the Paris Symposium.  He pointed out that tremendous 
energy, $10^{60}$ ergs, resided in extragalactic radio sources.  This was an unprecedented
amount for the time, and the paper was influential in forcing people to think about non-stellar
sources of energy in galaxies, that is, what we now call activity in galaxies.  

To continue with the topic of milestone years, we can also note that the evolution of the
quasar population was discovered by Schmidt in 1968, thirty years ago, a discovery that 
was an essential first step to showing that the characteristic time-scale for quasar activity
is quite short in cosmological terms.

\section{Definitions}

Schmidt's classic definition is that quasars are star-like objects 
of large redshift.  More
quantitatively, quasars are generally considered to have 
$z > 0.1$ and $M_B < -23~(H_0=50)$ mag (see Schmidt and Green 1983).
Traditionally, i.e. at resolutions of $1- 2$ arcsec, they were considered as 
being star-like, a description that is intertwined with the redshift and absolute
magnitude limits just given.  We now know that better spatial resolution often
yields evidence of a host galaxy.  Other key properties of quasars are that they
have broad emission lines\footnote{For this article, BL Lac objects will
be considered as a separate class, although they are members of the
AGN family.} in their spectra and that they can emit continuum radiation
across the electromagnetic spectrum from $\gamma$-rays to radio waves, with
ultraviolet and X-ray emission usually being very prominent.  Also, quasars show
variability on time scales of days to years.

How well can we explain all these properties?  Although it is generally
believed that the picture of an accretion disk surrounding 
a black hole is correct,
agreement between current models and observations is distressingly poor
in many cases.

Operationally, it is important to be aware of the effect of the 
apparent size and luminosity limit in the definition of quasars on 
modern surveys.  For example, as the angular resolution of surveys 
improves to 1 arcsec or better and the depth of surveys increases
either because of the use of larger aperture telescopes or longer
exposures, the host galaxies of quasars will be increasingly visible.
In such cases, strict imposition of the ``star-like'' criterion for
quasars will exclude bona-fide AGNs.  

Similarly, we now know that high-luminosity Seyfert galaxies can overlap
in absolute luminosity with low-luminosity quasars, and analyses of
surveys must allow for this.  For example, deep surveys for quasars
with good angular resolution will find both quasars and AGNs.  
It is important for the determination of the evolution of the entire
AGN population that surveys consider what classes of objects they
are including and perhaps rejecting.

\section{Recent Results}

\subsection{Host Galaxies}

The Hubble Space Telescope (HST) has provided critical new information on
the nature of the host galaxies in which quasars reside and about the
nature of quasar environments.  The excellent image quality of the
repaired telescope gives the best combination of angular
resolution and light gathering power yet applied to quasars.  Here 
I report on two papers, which of course build on previous ground-based
work.

Bahcall et al. (1997) presented results with the Wide-Field Camera of HST 
for 20 luminous quasars with $z < 0.3$.  For the host galaxies, they
found that 2 were as bright as the brightest cluster galaxies, 10 were
like normal elliptical galaxies, 3 were normal spirals, 3 were complex,
interacting systems, and in 2 cases there was faint nebulosity surrounding
the quasar.  For the radio-quiet quasars, 7 occurred in elliptical galaxies
and 3 in spirals.  For the 6 radio-loud quasars, 3 to 5 of them were
in elliptical galaxies.  On average, the host galaxies were 2.2 magnitudes
brighter than normal field galaxies.  In 8 cases, they detected companion
galaxies within a projected distance of 10 kpc from the quasar nucleus.
The interactions, presence of companions, and higher density of galaxies
seen around quasars suggest that interactions are important to quasar
activity.

Boyce et al. (1998) used HST in a complementary study of 14 low-redshift
quasars.  They find that 9 occur in elliptical galaxies (all 6 of
the radio-loud quasars and 3 radio-quiet objects); 2 radio-quiet quasars
are in disk galaxies, and the other 3, which are radio-quiet, ultraluminous
IR objects, occur in violently interacting systems.  The average luminosity
of the quasar host galaxies is 0.8 magnitudes brighter than $L^*$,
while the radio-loud objects are 0.7 magnitudes brighter than the
radio-quiet ones.

It is evident, as Bahcall et al. point out, that the hosts and 
environments of quasars are complex, and that the previous ideas 
about radio-quiet quasars residing in spiral galaxies
and radio-loud quasars in ellipticals may not
hold up.  However, it is perhaps more important to realize that
the HST observations provide powerful support for the concept of
quasars residing at the centers of galaxies and that galaxy
interactions play an important role in quasar activity.

\subsection{The Hubble Deep Field}

The Hubble Deep Field (HDF) has given us unprecedented new views of distant
galaxies.  In combination with spectroscopic observations with the
Keck Telescopes, studies of galaxies are now well advanced at $z > 3$,
redshifts that were unattainable previously.  Consequently, we now
have the opportunity to study directly the relationship of galaxies
and quasars at and beyond the redshift of peak quasar activity.  

Recently Conti et al. (1998) have carried out a detailed search for
compact quasars and AGNs in the HDF to $V_{606} = 27$ mag to study
their presence and behavior at luminosities corresponding to
AGNs in the nearby universe.  Although the HDF contains more than
3000 galaxies, Conti et al. found an upper limit of 20 for the number of
quasar candidates.  Based on spectroscopic observations to date, the
actual number may be much smaller, even close to 0.  However, because
of the great depth of the HDF exposures and the $\sim 0.1$ arcsec
image quality, it is possible that any AGNs in the HDF are spatially
resolved, and the next step is to develop sensitive techniques to 
detect AGN within faint, resolved galaxies in the HDF.  A complication
is that many of the distant galaxies being found by HST and Keck are
undergoing intense star formation, which gives them colors similar
to those of many quasars.  Jarvis and MacAlpine (1998) report identification
of 12 resolved objects harboring candidate AGN.  The crucial next step
will be to confirm the nature of the candidates with follow-up spectroscopy,
a very difficult task because of their faintness.

\subsection{Evolution of the Luminosity Function}

One of the most striking observed features of quasars is the
evolution of their luminosity function.  The space density of
luminous quasars increases by a factor of $\sim$1000 between
the present epoch and redshift $2-3$ and then falls steeply 
toward higher redshifts (Warren, Hewett, \& Osmer 1994, WHO;
Schmidt, Schneider, \& Gunn 1995, SSG; Kennefick, Djorgovski,
\& de Carvalho 1995).  A straightforward explanation
of this behavior is that we are seeing back to the epoch of peak
quasar activity, an epoch that presumably has to do with the
formation of black holes at the centers of galaxies and the time
of significant fueling of the quasar activity via the infall of
material to the center.

However, a persistent question about the nature of the peak is
whether it is affected significantly by dust absorption along
the line of sight.  If so, there could be an important population
of quasars at high redshift that are hidden at optical/UV wavelengths,
indicating that the epoch of peak activity was even earlier.  There
is no doubt that some quasars are highly reddened; the basic question
is how many.

One way to answer this question is to use samples of radio-selected
quasars with complete optical identifications.  Dust is transparent
to radio radiation, and so samples with complete optical identifications
provide an excellent test, as long as the ratio of radio quasars to
the total number of quasars does not change significantly with epoch.

Hook, Shaver, and McMahon (1998) have carried out just such a program
and find that the evolution of quasars in their sample is remarkably
similar to that found by WHO and SSG.  This suggests that dust is
not the cause of the apparent decline in activity at $z > 3$.  Similarly,
Benn et al. 1998 used IR observations in the K band of radio-selected
quasars and found no evidence for a large population of reddened and
dust-absorbed quasars.  These results are in contrast to those of
Masci (1998), who does claim evidence for a population of reddened
objects.

The ultra-deep ROSAT survey of Hasinger et al. (Hasinger 1998) has yielded
important new X-ray results.  The good positional accuracies of
the survey show that most of the sources are quasars/AGNs and narrow 
emission-line galaxies are only a small fraction, in contrast
with some previous work.  Their new
determination of the X-ray luminosity function is not consistent with
pure luminosity evolution but can be fit by pure density evolution from
$z=0$ to $z \approx 2$.
Their results suggest that black holes should be 
common in massive galaxies at the present epoch, as discussed in more detail below.   


\section{The News}

The astro-ph electronic preprint archive has had a large impact
on our field by greatly increasing the accessibility of preprints and
making them instantly available around the world.  It also provides
a convenient way of tracking the latest developments.  Here I mention
a few highlights gleaned from postings to astro-ph in the last year and
from other sources.

{\bf{The Most Luminous}}. Irwin et al. (1998) reported the discovery
of APM $0279+5255$, a broad-absorption line quasar with $z=3.87$ and
$R=15.2$ mag.  The object is coincident with an IRAS FSC source, and
the estimated luminosity is $\approx 5 \times 10^{15} L_{\odot}$, making it
the intrinsically most luminous object known.  There is evidence that
the source is gravitationally lensed, which amplifies the true emitted
luminosity.

{\bf{The Most Distant}}.  Weymann et al. (1998) find from Keck spectroscopy
an emission line in the galaxy HDF4-473.0 that, if identified with
Ly$\alpha$, yields a redshift of $z=5.60$.  The galaxy is in the Hubble
Deep Field and is the most distant object with slit spectroscopy that
has yet been identified.  It is not a quasar or AGN, and the absence of
quasars with $z > 4.9$, despite continuing surveys for them,
is beginning to appear 
significant in view of the increasing number of confirmed and candidate
galaxies with $z>5$. 

{\bf{The Smallest}}.  Kedziora-Chudczer et al. (1998) observed significant
radio variability on timescales less than an hour in the radio quasar
PKS $0405-385$, which would make it the smallest extragalactic source
observed.  They attribute the variation to interstellar scintillation
of a source with an angular size smaller than 5 microarcsec.  The inferred
brightness temperature is well above the inverse Compton limit.  If 
interpreted as steady relativistic beaming, the Lorentz factor would
be 1000.

{\bf{The first FIRST gravitational lens}}.  Schechter et al. (1998) found
that the quasar FBQ $0951+2635$, with $V=16.9$ mag and $z=1.24$, from
the FIRST radio survey, is a gravitational lens with two images separated
by 1.1 arcsec.

{\bf{Update to the Ver\'{o}n-Cetty and Ver\'{o}n Catalog}}.  
Ver\'{o}n-Cetty and Ver\'{o}n released the 8th edition of their catalog
during the year.  It contains entries for 11,358 quasars, 357 BL Lac objects,
and 3334 AGNs and is available electronically at 
http://obshpz.obs-hp.fr/www/catalogues/veron2\_8.html.  Such catalogs 
continue to be an vital resource for the community, especially as new
surveys yield so many new quasars and AGNs.  Also, the electronic
availability of the catalogue makes it even more accessible and valuable
than it was previously.

\section{Some Current Research Problems}

Here I call attention to some current research problems that need 
further work.  Their eventual solution should improve our understanding
of quasars and AGNs in important ways.

{\bf{The Disagreement between Observations and Predictions for Accretion
Disks}}.  Koratkar (1997) points out that observations do not confirm
most predictions of accretion disk models.  For example, the Zheng et
al. (1997) composite spectrum for ultraviolet wavelengths does not 
match predictions, and soft X-ray fluxes are observed to be too flat.
Fewer Lyman edges are observed than predicted.
Polarization is not seen either, which seems to rule out scattering
as a way of smoothing the Lyman edges.  An additional theoretical
question is how the radiation from the accretion disk couples with
that of the hot (X-ray) corona.  It is important to resolve
these issues if we are to have confidence in this basic
part of our concept for quasars and AGNs.

{\bf{What powers Ultra-luminous IRAS galaxies?}}  Observations by
Genzel et al. (1998) indicate that massive stars predominate in 
70--80\% of the cases, with AGNs dominating in the others.  At
least half of the systems probably have both an AGN and a 
circum-nuclear ring of starburst activity.  They see no clear
trend for the AGN component to dominate in the most compact
and presumably most advanced mergers.  

{\bf{Do all galaxies have massive black holes?}}  van der Marel
(1997) notes that available data appear consistent with most
galaxies having black holes, whose mass roughly correlates with
the luminosity of the spheroid (cf. Magorrian et al. 1998).  
The black holes could have formed
in or prior to a quasar phase and grown via mass accretion.  Some
of the implications of this work are discussed below under the
theory section.

{\bf{Is the broad Fe K$\alpha$ line produced directly near
a black hole? How well do we understand the origin of X-ray emission
in general?}}  Observations of the broad Fe K$\alpha$ line
in AGNs are widely interpreted as arising in the inner part
of accretion disks around black holes and therefore providing
both confirmation of the presence of black holes as well as
direct information about conditions in the disks.  However,
Weaver and Yaqoob (1998) have raised questions about whether
the emission in fact does occur so close to the centers of AGNs.
More generally, intensive monitoring of NGC 7469 in X-rays and the
ultraviolet by Nandra et al. (1998) provides strong constraints on
quasar models.  The data are not consistent with the UV emission being
reprocessed by gas absorbing X-rays nor with the X-rays arising
from Compton upscattering of the UV radiation.

These are just some examples of research problems in need of
solution for us both to have confidence in our general picture
of quasars and AGNs being powered by accretion onto massive
black holes and to develop a quantitative understanding that
explains the major observed features of these objects.

\section{Theory}

In addition to the above types of problems, considerable research
is directed to basic questions such as, Do we understand how quasars 
form and evolve?  Can we connect
theories of galaxy and black hole formation with the observations
of quasars at high redshift and the incidence of black holes in
galaxies at low redshift?  Here I mention briefly some recent theoretical
work that demonstrates progress in our understanding of quasars and
ties in with present and future observational work.

Haiman, Madau, and Loeb (1998) point out that the scarcity of quasars
at $z > 3.5$ in the Hubble Deep Field implies that the formation of
quasars in halos with circular velocities less than 50 km/s is
suppressed (on the assumption that black holes form with constant
efficiency in cold dark matter halos).  They note that the Next 
Generation Space Telescope should be able to detect the epoch of
formation of the earliest quasars.

Cavaliere and Vittorini (1998) note that the observed form for the evolution
of the space density of quasars can be understood at early times when
cosmology and the processes of structure formation provide material for
accretion onto central black holes as galaxies assemble.  Quasars then
turn off at later times because interaction with companions cause the
accretion to diminish.

Haehnelt, Natarajan, and Rees (1998) show that the peak of quasar
activity occurs at the same time as the first deep potential wells
form.  The Press-Schechter approach provides a way to estimate the
space density of dark matter halos.  But the space density of $z=3$
quasars is less than 1\% that of star-forming galaxies, which implies
the quasar lifetime is much less than a Hubble time.  For an assumed
relation between quasar luminosity and timescale and the Eddington
limit, it is possible to connect the observed quasar luminosity density
with dark matter halos and the numbers of black holes in nearby galaxies.
The apparently large number of local galaxies with black holes implies
that accretion processes for quasars are inefficient in producing blue
light.

\section{Future Directions and Possibilities}

The research problems and theoretical ideas described in this article
are already open to observational study and testing with 8-10-m class 
telescopes and the Hubble Space Telescope, as we have discussed
in the case of studies of quasar host galaxies, high-redshift galaxies,
and black holes in galaxies.  As the capabilities of the large 
ground-based telescopes improve (via infrared optimization and
adaptive optics, for example), 
and when the Next
Generation Space Telescope is completed, we will be able to study
directly the relation of AGNs and galaxies over virtually the
entire range of their evolutionary history.  Similarly, the X-ray
observatories AXAF and XMM will offer very significant new capabilities
for the study of both the nature of quasars and AGNs and their evolution.

In the meantime, large-area, ground-based surveys such as the Sloan Digital
Sky Survey\footnote{www.sdss.org}
and the 2dF\footnote{msowww.anu.edu.au/$\sim$rsmith/QSO\_Survey/qso\_surv.html}
survey will increase the number of known quasars
by more than an order of magnitude.  We may expect that the combination of 
the new samples, the new observatories, and continued theoretical advances
will answer many of the questions raised here. 

\acknowledgements I thank Brad Peterson and David Weinberg 
for comments and suggestions on a 
first draft of this article.  I am grateful to the Organizing Committee
and the National Science Foundation (via grant AST-9529324) for financial support.

\end{document}